# Self-compression to sub-3-cycle duration of mid-infrared optical pulses in bulk

Michaël Hemmer[1], Matthias Baudisch[1], Alexandre Thai[1], Arnaud Couairon[2], Jens Biegert[1,3]

The generation of few-cycle pulses with controlled waveforms in the mid-infrared spectral region is a long-standing challenge but is expected, to enable a new generation of high-field physics experiments, uncovering intricate physical phenomena. Successful generation of such optical pulses is limited by the tremendous spectral width – exceeding 1000 nm – required to withstand few-cycle pulses in the mid-IR correlated with the need to tightly control the spectral phase over such a broad bandwidth. Here, we present the first demonstration of sub-3 cycle optical pulses at 3.1 μm central wavelength using for the first time self-compression in the anomalous dispersion regime in bulk material. The pulses emerging from this compact and efficient self-compression setup could be focused to intensities exceeding $10^{14}$ W/cm$^2$, a suitable range for high field physics experiments. Our experimental findings are corroborated by numerical simulations using a 3D nonlinear propagation code, therefore providing theoretical insight on the processes involved.

Intense few-cycle pulses are difficult to obtain directly from a laser system due to limitations in amplification bandwidth, spectral reshaping and dispersion management. A solution to this problem is nonlinear propagation in gaseous media in capillaries [1] or free-space (filamentation) [2,3] and subsequent compression with chirped mirrors [4]. These techniques can reduce pulse durations from the typical 30 fs at 800 nm to below two-cycles, but, depending on implementation, efficiencies and pulse durations across the output beam may vary [5]. A lesser-known method to compress transform-limited pulses is based on group-velocity mismatch and fast amplitude modulation during three-wave-mixing in a nonlinear crystal [6]. The technique was demonstrated for Nd:YAG wavelength and allowed compression from 10 ps duration to 310 fs while doubling the input frequency [7]. Self-compression to the few-cycle regime, i.e. without the need for dispersion compensation, was predicted [8] and achieved at 800 nm and 1500 nm wavelengths [9], but is limited in pulse energy due to rapid and chaotic pulse splitting [10,11].

The current upsurge in activity to generate intense few-cycle pulses in the mid-IR motivates revisiting nonlinear propagation and pulse compression in those regimes. A limitation for applying the above-mentioned techniques is the typically insufficient pulse energy from mid-IR laser systems, which makes using gaseous media for nonlinear propagation challenging. The combination of low pulse energy and propagation in the anomalous dispersion regime is however suited to investigate self-compression in bulk media due to the much higher nonlinearities.

In this Letter, we demonstrate stable and efficient self-compression of mid-IR pulses in bulk material via filamentation in the anomalous dispersion regime. The spectro-temporal properties of the pulses resulting from this highly nonlinear interaction have been investigated and revealed durations as short as 3 optical cycles with energy throughput as high as 80% (65% without correcting for Fresnel losses due to uncoated optics). The mid-IR seed pulses used for the experiment were generated by the home-built optical parametric chirped pulse amplifier (OPCPA) system described in Ref. [12]. We used 3 μJ of the available 12.5 μJ, 70-fs pulses at 3100 nm center wavelength; the output spatial profile is excellent and power stability is typically better than 1% rms over 30 min. The laser system operates at 160 kHz repetition rate. Figure 1 shows the setup, consisting of an uncoated 10 cm focal length CaF$_2$ lens and a 3-mm thick, uncoated, Yttrium Aluminum garnet (YAG) plate which is moved 6.5 mm along the beam across the focal plane in steps of 0.5 mm so as to experience different peak intensities. The emerging beam was collimated with a silver-coated, 15 cm focal length mirror and the pulse characterized with frequency-resolved optical gating (FROG). Our FROG device was specifically designed to handle ultrabroad bandwidths in the mid-IR spectral region [13] and allowed lifting typical time axis direction ambiguity; mid-IR spectra were measured using a Fourier transform infrared spectrometer with a liquid nitrogen cooled MCT detector.

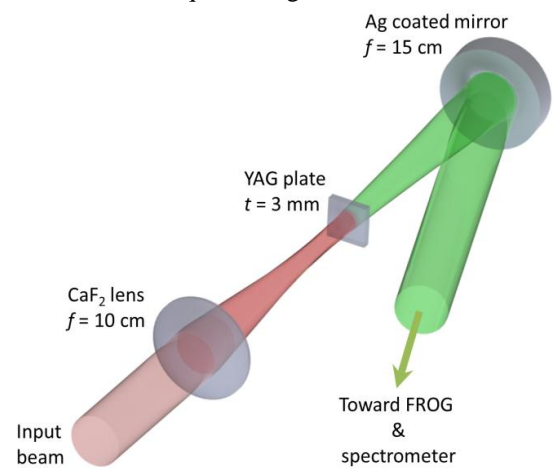

**Figure 1| Layout of the self-compression setup**. The input beam is focused by a 10 cm focal length CaF2 lens into a 3 mm-thick YAG plate. The beam emerging from the plate is collimated with a 15 cm focal length silver coated mirror and directed toward diagnostic tools. In this study, the YAG plate was moved along the beam propagation axis to investigate the behavior of the self-compression process. The Fresnel losses of the uncoated optics amount to 15%.

The theoretical model used in this work was described elsewhere [14] and computes nonlinear propagation of the frequency components of the mid-infrared pulse in the YAG plate using a radially symmetric carrier resolving model in the form of the forward Maxwell equation coupled with an evolution equation for the electron-hole and electron-ion plasma. In this model, diffraction is described in the paraxial approximation, dispersion of YAG is described by a Sellmeier relation [15], self-steepening and nonlinear chromatic dispersion are included as well as optical field ionization rates [16] and avalanche ionization via the Drude model.

We have previously investigated and reported [16] the spatio-spectral behavior of supercontinuum generation in a setting similar to that shown in Fig. 1 and simulations had predicted that self-compression was expected to occur in a narrow region of the parameter space defined by the input energy, seed pulse duration, focal beam size and YAG plate thickness. In order to experimentally map such a wide parameter space, a few parameters were set based on preliminary simulations: the pulse duration was set to the shortest duration directly achievable at the output of the OPCPA system (70 fs equivalent to 7-cycle duration), the YAG plate thickness was set to 3 mm, the focal length of the focusing lens was set to 10 cm. Focal lengths of 5 and 7.5 cm have also been tested and led to lesser performances. Both the impinging energy and the location of the YAG plate in the focus were left as free parameters: tuning of the YAG plate position allowed tuning of the peak intensity and of the length of filamentary propagation in the YAG plate while modifying the energy allowed fine adjustment of the peak intensity.

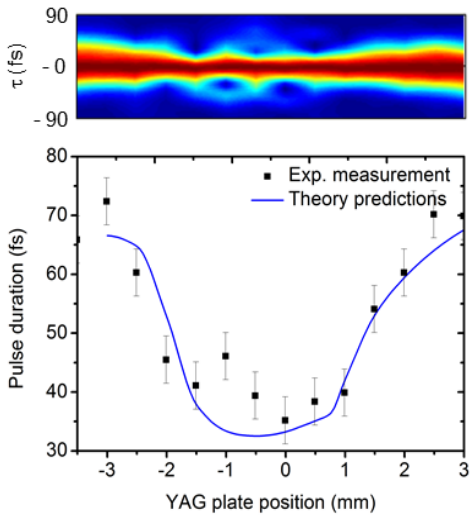

**Figure 2|** **Measured and simulated evolution of the pulse duration as the seed pulse undergoes self-compression while the YAG plate is scanned through focus.** (Top) Retrieved experimental temporal profile of the pulse intensity as a function of the YAG plate position; (bottom) experimentally measured (black dots) and simulated (blue line) evolution of the pulse duration as a function of YAG plate position through focus.

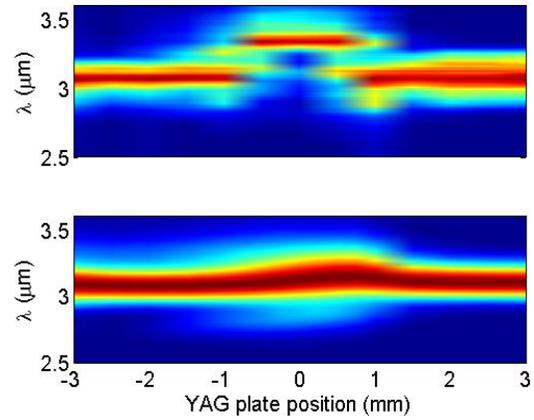

**Figure 3|** **Measured and simulated evolution of the spectrum as the seed pulse undergoes temporal self-compression.** (Top) Measured evolution of the spectral intensity profile as the YAG plate is scanned through focus; (bottom) simulated evolution of the spectral intensity profile as the YAG plate is scanned through focus.

In order to map the leftover parameter space and demonstrate the experimental feasibility of self-compression, the YAG plate was scanned through the focus of the 10 cm focal length lens in steps of 0.5 mm over a 6.5 mm distance. For each position of the YAG plate, a FROG trace was recorded along with the corresponding spectrum and the corresponding pulse duration was retrieved. The experimental data for the temporal domain are compiled in Fig. 2 while those for the spectral domain are gathered in Fig. 3. The simulation results in both temporal and spectral domains are added to the corresponding figures. The origin of the position of the YAG plate was chosen to be the location of the focus – determined independently by measuring the beam size through the focus. The YAG plate position indicated on Fig. 2 is the distance between the front of the YAG plate and that of the focus. Negative values are located closer to the focusing lens. The evolution of the seed pulse intensity as the YAG plate is scanned through focus shows qualitatively that self-compression does occur and that no pulse splitting is observed even though the onsets of pulse splitting is observable for positions around -1 mm (Fig. 2 (top)). The highest intensity at the entrance face of the YAG plate was evaluated to be ~2 TW/cm$^2$ at the focus, making valid the assumption that no significant air ionization takes place. The hypothesis that optimum self-compression occurs at the onset of pulse splitting was confirmed by placing the YAG plate in the region comprised between the abscissa [-1,1] mm and increasing the pulse energy which readily resulted in pulse splitting. It was found that when sending the full energy available from the driving system (12.5 μJ) – corresponding to peak intensities at focus of ~8 TW/cm$^2$ – the plate could be scanned through the focus without any damage; measuring the temporal profile of the pulses emerging from the YAG plate at this high energy showed multiple splitting of the seed pulse in the region around the abscissa [-1,1] mm while partial self-compression could be recovered if the plate was placed beyond the focus (i.e. for positive abscissa). The well-defined retrieved intensity profiles showed in Fig. 2 (top) allowed extracting FWHM duration of the pulse without ambiguity. The extracted FWHM durations are shown in Fig. 2 (bottom) and are overlaid with the durations expected from numerical simulations. The striking agreement was obtained for a Kerr coefficient $n_2 = 6 \times 10^{-16}$ cm$^2$/W with 50 % contribution of a delayed Raman-Kerr effect, a collision time of 20 fs in the Drude model and an optical field ionization rate W(I) lower than its multi-photon asymptote $\sigma_{17}I^{17}$ with coefficient $\sigma_{17} = 2 \times 10^{-203}$ cm$^{34}$/sW$^{17}$. The intensity distribution was radially averaged over a 50 μm radius that is always at least twice as larger as the filament radius. Pulse durations calculated by radially averaging the beam profile over 25 μm are shorter by only a few percent, indicating that the high intensity filament and its periphery were homogeneously shortened. If the filament core were much shorter, the weight of the filament periphery would ruin the overall pulse compression. Simulations reveal the mechanism for optimal compression: while the beam diameter is maintained approximately constant throughout the YAG plate, the pulse shrinks due to the strong anomalous dispersion ($k_0^{''} = -4082$ fs$^2$/cm$^1$) and to steepening of the pulse tail by the self-generated electron-hole plasma.

The spectral behavior was simultaneously investigated and revealed a dramatic spectral broadening as the pulse experienced maximum self-compression accompanied by a large red-shift of the peak spectral intensity. The seed spectrum was found to broaden from ~600 nm width at 1/e$^2$ of the peak intensity outside the focus to almost 1000 nm at 1/e$^2$ of the peak intensity in the focal region while the center wavelength shifted by almost 300 nm, from 3100 nm to 3400 nm. Investigating the evolution of the spectral phase through focus revealed that the self-compressed pulses exhibited residual uncompensated phase attributed to self-phase modulation, suggesting the potential for further compression. Even though the simulation and experimental traces show qualitative agreement – spectral broadening and red-shifting – the magnitude of the red-shift is larger experimentally than theoretically expected. Several attempts have been made to investigate this discrepancy – the main challenge laying in the absence of data for the Raman-Kerr response of our crystal – particularly by adjusting the Raman contribution. Numerically increasing the Raman-Kerr contribution from 20% to 90% of the total Kerr index in the nonlinear propagation or modifying step by step the frequency shift from 0.05 to 0.3 PHz in the Raman response did result in a moderate increase in the magnitude of the observed red-shift, while the pulse duration retrieved by simulations was not significantly affected by these changes. The discrepancy indicates that the observed redshift may not be related to the Raman-Kerr response of the crystal and is not prevalent for reaching the optimal pulse compression.

Based on the previous systematic scanning and the simulation results, the optimum self-compression conditions were found and the optimally self-compressed pulses characterized (Fig. 4). In the optimal setting the temporal and spectral intensity profile of the seed pulses were carefully measured (Fig. 4 (top)). The retrieved spectral intensity profile of the seed pulses is shown in Fig. 4 (top-right) and shows good agreement with the measured spectrum confirming the accuracy of the retrieval (0.5% error). The optimum self-compressed pulse was obtained when the YAG plate was placed exactly at the focus of the 10 cm focal length lens with an input energy of 3 μJ. The measured spectrum for the optimally self-compressed pulse shows an extend of over 1000 nm and the retrieved spectrum from the FROG measurement is in good agreement with the measured spectrum confirming again the consistency of the measurement. The retrieved FWHM pulse duration is 32 fs which corresponds to ~3-cycles of electric field at 3100 nm. The spatial profile of the pulse was measured using a knife-edge technique and showed that the process left the seed spatial profile undisturbed. In order to investigate the viability of this technique as a source of few-cycle pulses in the mid-infrared – as an equivalent to the well-established gas-filled hollow-core fiber based techniques used in the near-infrared – the throughput of this self-compression setup was investigated (Fig. 5). It was experimentally found that the throughput of the setup when the uncoated YAG plate was placed largely outside the focus – i.e. when minimal nonlinear effect took place – was 85%, the 15% losses closely matching that of Fresnel losses expected from a plate with an index of refraction with that of YAG. Upon sliding the YAG plate to the position where optimum self-compression occurred, the throughput of the system was reduced to 65% (Fig. 5 – top). Using the same numerical setting conditions as previously, we numerically investigated the source of the extra losses inherent to the self-compression and found that those losses originate from ionization and plasma absorption.

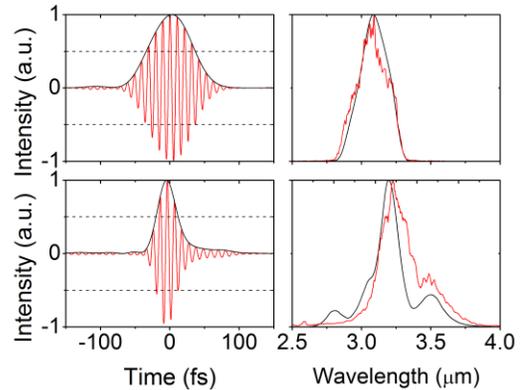

**Figure 4| Measurement of spectro-temporal profile of the self-compressed pulses.** (Top left) Retrieved normalized 7-cycle duration intensity profile of the seed pulses along with a calculated oscillating field with center wavelength $\lambda_o$ = 3.1 μm; (top right) measured (red line) and retrieved (black line) spectra of the seed pulses; (bottom left) retrieved normalized sub-3-cycle duration intensity profile of the self-compressed pulses along with a calculated oscillating field with center wavelength $\lambda_o$ = 3.1 μm; (bottom right) measured (red line) and retrieved (black line) spectra of the self-compressed pulses.

Coefficients of the model are not known with great accuracy but a numerical variation of the collision time from 1 to 20 fs and of the optical field ionization rates over 3 decades around the Keldysh values did not qualitatively affect the compression process or the behavior of the transmission as functions of the position of the YAG plate. Quantitatively, ionization and plasma absorption highly depend on model coefficients but the filament intensity self-adjusts so as to maintain overall losses at the same level, mainly determined by the plate thickness and focusing geometry. While the ionization and plasma absorption losses are unavoidable, applying a suitable dielectric coating to the YAG plate could enable a global throughput of 80%, a figure largely exceeding that of typical nonlinear compressor used at near-infrared wavelengths. Finally, the shot-to-shot stability of the self-compressed pulses was measured to be 0.8% rms over 10 minutes. Long term drifts were not observed and the experiment could be repeated over days.

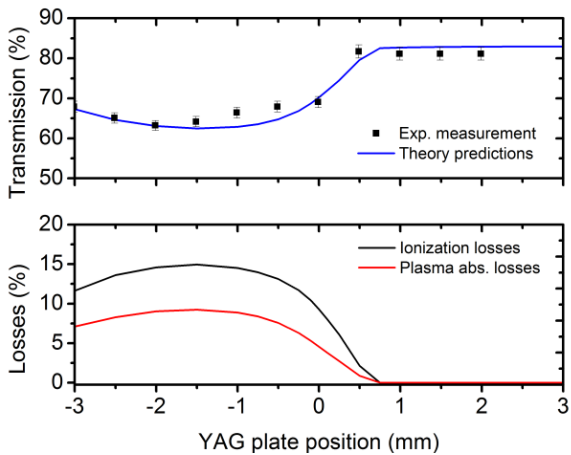

**Figure 5| Measured and simulated transmission of the self-compression setup and contribution of ionization and plasma absorption losses as the YAG plate is scanned through focus.** Simulated (blue line – top) and measured (squares – top) transmission reveal the high efficiency of the process and suggest potential throughput of 80% if anti-reflection coating is applied to the YAG plate. The proportion of losses inherent to the self-compression attributed to ionization (black curve – bottom) and plasma absorption (red curve – bottom) have been identified.

In conclusion, we demonstrated for the first time that self-compression to the few-cycle duration by nonlinear propagation in the anomalous dispersion regime in condensed media is feasible and stable. We demonstrate the generation of sub-3 cycle pulses in the mid-infrared spectral region with sufficient energy (1.9 µJ) to reach peak intensities in the $10^{14}$ W/cm$^2$ range and outstanding pulse-to-pulse stability (0.8% rms) at 160 kHz repetition rate, three critical features to enable strong-field physics experiments. The conditions for successful self-compression have been experimentally and theoretically investigated and the conjunction of both of these approaches enabled the generation of record few-cycle pulse durations.

**Acknowledgements**

We acknowledge support from MINISTERIO DE ECONOMÍA Y COMPETITIVIDAD through its Consolider Program Science (SAUUL CSD 2007-00013), through Plan Nacional (FIS2011-30465-C02-01), the Catalan Agencia de Gestio d'Ajuts Universitaris i de Recerca (AGAUR) with SGR 2009-2013, Fundacio Cellex Barcelona, and funding from LASERLAB-EUROPE, grant agreement 228334.

**Author contributions**

M.B., A.T. and M.H. set up and performed the experimental work. M.H. and J.B. analyzed the experimental data and interacted with A.C. who performed the numerical simulations. M.H., A.C. and J.B. wrote up the paper.